\documentclass[11pt]{amsart}
\usepackage[T1]{fontenc}
\usepackage{amssymb,amsthm,mathtools}
\usepackage{newunicodechar}
\usepackage{xcolor}
\definecolor{linkblue}{RGB}{0,82,204}
\usepackage[colorlinks,linkcolor={red!90!blue},citecolor=black,urlcolor=linkblue]{hyperref}
\usepackage{booktabs,longtable,array}
\usepackage{listings}

\newunicodechar{ℕ}{\ensuremath{\mathbb{N}}}
\newunicodechar{ℙ}{\ensuremath{\mathbb{P}}}
\newunicodechar{ℤ}{\ensuremath{\mathbb{Z}}}
\newunicodechar{→}{\ensuremath{\to}}
\newunicodechar{ₑ}{\ensuremath{{}_e}}

\newcommand{\lcode}[1]{\texttt{#1}}
\newcommand{\I}{\mathbb{I}}
\newcommand{\F}{\mathcal{F}}

\theoremstyle{plain}
\newtheorem*{honesty}{Remark}

\title[The set of primes is supernatural: a Lean formalization]{The set of primes is supernatural: a Lean formalization of the statement of the conjecture}
\author{Arnaud Mayeux}
\address{University of Wisconsin--Madison, Madison, WI, USA}
\email{mayeux@wisc.edu}

\begin{document}
\begin{abstract}
The paper \emph{Conjecture: the set of prime numbers is supernatural} conjectures that
no non-constant function built
from the identity and constants by finitely many pointwise additions, multiplications,
and exponentiations maps every positive integer to a prime. We give a complete Lean~4
formalization of that paper over Mathlib: every definition, example, remark, numbered
result, and experimental table row has a machine-checked counterpart, with no
\lcode{sorry}. The conjecture and similar generalizations are stated exactly, as named
open problems. So stated, the conjecture becomes a precise target:
an automated reasoning system can now attempt a kernel-checked proof.
\end{abstract}

\maketitle
\setcounter{tocdepth}{1}
\tableofcontents

\section{Introduction}

The paper \cite{Mayeux} calls a function on the positive integers \emph{natural} if it is
built from the identity and constants by finitely many applications of $+,\times,\wedge$,
and states a single precise conjecture about the set of primes $\mathbb{P}$: no
non-constant natural function sends every positive integer to a prime. To formulate this,
it introduces \emph{elevation structures}, axiomatizing the compatibilities of addition,
multiplication, and exponentiation; around the conjecture it develops a compact theory
and reports computer experiments. The conjecture is expected in \cite{Mayeux} ``to be a
hard challenge for any kind of intelligence,'' human or artificial.

The goal of the present paper is threefold.
\begin{enumerate}
\item To formalize fully the paper \cite{Mayeux}, a contribution to the programme
  described in \cite{Benchmark}.
\item To extend, both mathematically and formally, some of \cite{Mayeux}'s remarks,
  namely the extensions sketched in \cite[Remark~3.4]{Mayeux}: Knuth arrows and the factorial, combined
  here into a single enlarged class that also admits truncated subtraction.
\item To formalize the \emph{statement} of the conjecture from \cite{Mayeux}, as a named,
  unproved Lean proposition: a precise target whose eventual proof, human or automated,
  can be checked by the kernel.
\end{enumerate}
Let us here ask the following question: which of the two benchmarks, formalizing all of
indexed mathematics or settling this particular conjecture, will be solved first?

The present paper is a companion to a Lean~4 \cite{Lean4} formalization of that material on
top of Mathlib \cite{mathlib}, kept in a single file \lcode{SPCL.lean} (a standalone
project whose only dependency is Mathlib; Lean toolchain \lcode{v4.32.2}). The source is
openly available at
\begin{center}
\url{https://github.com/rndmx/SPCL}
\end{center}
The file is about $1200$ lines, all of it compilable code: it carries no comments; the
exposition and the correspondence with the printed paper live entirely here. It builds,
\lcode{sorry}-free, in about half a minute (\S\ref{sec:filequick}).

\subsection*{Coverage}
The formalization is \emph{exhaustive} with respect to the printed paper: every
definition, example, remark, numbered result, and experimental table row of \cite{Mayeux}
has a Lean counterpart, treated in full, from the category structure and the word
construction (with its proved equivalence to the inductive encoding) to each individual
sample function and every table row. Section~\ref{sec:table} lists the complete
correspondence.

\subsection*{How to read this paper}
Each section below states the mathematics as in \cite{Mayeux} (statement numbers
``Definition~1.1'', ``Proposition~2.2'', etc.\ refer to the printed paper), then presents
the corresponding Lean declarations verbatim, with commentary on any distance between the
two, namely encoding choices, coercions, or proof-level differences. No Lean background is
assumed beyond the following: \lcode{ℕ+} (Mathlib's \lcode{PNat}) is the type of positive
integers, the paper's $\I$; a term of type \lcode{Prop} is a mathematical statement, and a
\lcode{theorem} is a statement together with a checked proof; \lcode{def C : Prop := ...}
\emph{names} a statement without asserting it; an \lcode{instance} registers a structure on
a type so that later declarations can use it silently. A reader coming from the code side
can conversely take the mathematical paragraphs as the specification each declaration is
answerable to.

\subsection*{Conventions}
Lean reserves $\wedge$ for logical conjunction, so the elevation operation is named
\lcode{elev}, with scoped infix notations \lcode{+ₑ}, \lcode{*ₑ}, \lcode{\^{}ₑ} for the
three operations of an abstract elevation structure. The paper's $\I$ is \lcode{ℕ+}
throughout; the coercion \lcode{(n : ℕ)} sends \lcode{n : ℕ+} to the corresponding natural
number, and exponentiation on \lcode{ℕ+} takes its exponent in \lcode{ℕ}, which is why
\lcode{f n \^{} (g n : ℕ)} is the pointwise $f(n)^{g(n)}$.

\section{The file at a glance}\label{sec:filequick}

\lcode{SPCL.lean} is a single file, following the order of the printed paper so that the
correspondence can be audited linearly: the theory of \cite{Mayeux} inside the
\lcode{SPCL} namespace, from elevation structures through the results of \cite[\S3]{Mayeux} and the
extension of \cite[Remark~3.4]{Mayeux}, ending with the closing question; then the
\lcode{Pratt} namespace with the primality certificates (\S\ref{sec:tables}); and finally
the $89$ experimental-table theorems of \cite[\S4]{Mayeux}, which are statements about
bare numerals and therefore live at top level.

The project is standard Lake: \lcode{lake build} checks everything (about half a minute);
any single claim can be inspected interactively with
\lcode{\#check} by the names used throughout this paper. The file imports all of Mathlib
for convenience. No \lcode{axiom} is declared anywhere; \lcode{\#print axioms} on any of
the theorems reports at most Lean's three standard classical axioms (propositional
extensionality, quotient soundness, choice), inherited through Mathlib. The named
conjectures are \lcode{def ... : Prop} and therefore contribute no assumptions at all.
The file contains no comments: every declaration is meant to be read side by side with
this paper, which is its sole documentation.

\section{Elevation structures and their category (\texorpdfstring{\cite[Definition~1.1, Example~1.2]{Mayeux}}{[3, Definition 1.1, Example 1.2]})}

\noindent\textbf{Definition~1.1 (\cite{Mayeux}).} An \emph{elevation structure} is a
$4$-tuple $(E,+,\times,\wedge)$ where $E$ is a set and $+,\times,\wedge : E\times E \to E$
are maps such that for all $a,b,c\in E$:
\[
\begin{aligned}
&\text{(i) } a+b=b+a, \qquad \text{(ii) } a+(b+c)=(a+b)+c,\\
&\text{(iii) } b\times a = a\times b, \qquad \text{(iv) } a\times(b\times c) = (a\times b)\times c,\\
&\text{(v) } a\times(b+c)=a\times b + a\times c,\\
&\text{(vi) } (a\wedge b)\times(a\wedge c) = a\wedge(b+c), \qquad \text{(vii) } (a\wedge b)\wedge c = a\wedge(b\times c).
\end{aligned}
\]
Crucially, $\wedge$ itself need be neither commutative nor associative: $(a\wedge b)\wedge c
\neq a\wedge(b\wedge c)$ in general, and the paper fixes the convention that
$a_1\wedge a_2\wedge\cdots\wedge a_n$ means $a_1\wedge(a_2\wedge(\cdots\wedge a_n))$, i.e.\
$\wedge$ associates to the right. A \emph{morphism} of elevation structures is a map
preserving $+,\times,\wedge$. \emph{We obtain a category.}

\medskip
\noindent\textbf{Formalization.} \cite[Definition~1.1]{Mayeux} becomes a type class, and a morphism a
structure bundling a map with the three compatibility laws:
\begin{lstlisting}
class ElevationStructure (E : Type*) where
  add : E → E → E
  mul : E → E → E
  elev : E → E → E
  add_comm : ∀ a b : E, add a b = add b a
  add_assoc : ∀ a b c : E, add a (add b c) = add (add a b) c
  mul_comm : ∀ a b : E, mul a b = mul b a
  mul_assoc : ∀ a b c : E, mul a (mul b c) = mul (mul a b) c
  mul_add : ∀ a b c : E, mul a (add b c) = add (mul a b) (mul a c)
  elev_add : ∀ a b c : E, mul (elev a b) (elev a c) = elev a (add b c)
  elev_mul : ∀ a b c : E, elev (elev a b) c = elev a (mul b c)

structure ElevationHom (E F : Type*) [ElevationStructure E] [ElevationStructure F] where
  toFun : E → F
  map_add' : ∀ a b, toFun (a +ₑ b) = toFun a +ₑ toFun b
  map_mul' : ∀ a b, toFun (a *ₑ b) = toFun a *ₑ toFun b
  map_elev' : ∀ a b, toFun (a ^ₑ b) = toFun a ^ₑ toFun b
\end{lstlisting}
The sentence ``we obtain a category'' is itself formalized: identity, composition, and the
three category laws, each of which holds definitionally (\lcode{rfl}):
\begin{lstlisting}
def ElevationHom.id (E : Type*) [ElevationStructure E] : ElevationHom E E where
  toFun := _root_.id
  map_add' _ _ := rfl
  map_mul' _ _ := rfl
  map_elev' _ _ := rfl

def ElevationHom.comp {E F G : Type*} [ElevationStructure E] [ElevationStructure F]
    [ElevationStructure G] (M : ElevationHom E F) (N : ElevationHom F G) :
    ElevationHom E G

theorem ElevationHom.id_comp (M : ElevationHom E F) : (ElevationHom.id E).comp M = M := rfl
theorem ElevationHom.comp_id (M : ElevationHom E F) : M.comp (ElevationHom.id F) = M := rfl
theorem ElevationHom.comp_assoc (M : ElevationHom E F) (N : ElevationHom F G)
    (P : ElevationHom G H) : (M.comp N).comp P = M.comp (N.comp P) := rfl
\end{lstlisting}
(Composition is written diagrammatically: \lcode{M.comp N} is ``$M$ then $N$''.) The
category is also registered as an actual instance of Mathlib's \lcode{Category} class, on
the type of bundled elevation structures:
\begin{lstlisting}
structure ElevCat where
  carrier : Type u
  [str : ElevationStructure carrier]

instance : CategoryTheory.Category ElevCat where
  Hom E F := ElevationHom E F
  id E := ElevationHom.id E
  comp M N := M.comp N
  id_comp _ := rfl
  comp_id _ := rfl
  assoc _ _ _ := rfl
\end{lstlisting}

\medskip
\noindent\textbf{Example~1.2 (\cite{Mayeux}).} $(\I,+,\times,\wedge)$, with $\wedge$
ordinary exponentiation, is an elevation structure, where $\I$ denotes the positive
integers. So is $(\F,+,\times,\wedge)$, $\F$ the set of functions $\I\to\I$, with all three
operations pointwise; for $a\in\I$ the evaluation map $E_a:\F\to\I$, $f\mapsto f(a)$, is a
morphism.

\medskip
\noindent\textbf{Formalization.} Both halves are instances, and each evaluation map is an
actual term of \lcode{ElevationHom}:
\begin{lstlisting}
instance : ElevationStructure ℕ+ where
  add := (· + ·)
  mul := (· * ·)
  elev := fun a b => a ^ (b : ℕ)
  -- ⋮ (the seven axioms: PNat's semigroup laws plus pow_add / pow_mul)

instance : ElevationStructure (ℕ+ → ℕ+) where
  add f g := fun n => f n + g n
  mul f g := fun n => f n * g n
  elev f g := fun n => f n ^ (g n : ℕ)
  -- ⋮ (the seven axioms, pointwise, from the ℕ+ instance)

def Eval (a : ℕ+) : ElevationHom (ℕ+ → ℕ+) ℕ+ where
  toFun f := f a
  map_add' _ _ := rfl
  map_mul' _ _ := rfl
  map_elev' _ _ := rfl
\end{lstlisting}
For the $\I$ instance the two elevation axioms reduce to \lcode{pow\_add} and
\lcode{pow\_mul}; morphismhood of $E_a$ is definitionally trivial (\lcode{rfl} closes all
three laws), since $\F$'s operations were \emph{defined} to be pointwise applications of
$\I$'s.

\section{Natural functions: the word construction and the inductive closure
(\texorpdfstring{\cite[Definitions~1.3, 1.4, 1.7]{Mayeux}}{[3, Definitions 1.3, 1.4, 1.7]})}\label{sec:words}

\noindent\textbf{Definitions~1.3--1.4 (\cite{Mayeux}).} Let $\F_s\subset\F$ consist of the
identity map together with all constant maps (the \emph{part of symbols}). For
$P\subseteq\F$, let $A_+(P)$ (resp.\ $A_\times(P)$, $A_\wedge(P)$) be $P$ together with
every $g+h$ (resp.\ $g\times h$, $g\wedge h$) for $g,h\in P$. Writing $\Sigma$ for the set
of finite words in the alphabet $\{A_+,A_\times,A_\wedge\}$, the set of \emph{natural
functions} is
\[
\F_{\mathrm{Natural}} \;=\; \bigcup_{\sigma\in\Sigma}\sigma(\F_s).
\]
Moreover $(\F_{\mathrm{Natural}},+,\times,\wedge)$ is itself an elevation structure.

\medskip
\noindent\textbf{Definition~1.7 (\cite{Mayeux}).} The \emph{length} of a natural function
$f$ is the least length of a word $\sigma\in\Sigma$ with $f\in\sigma(\F_s)$.

\medskip
\noindent\textbf{Formalization, first encoding: the word machinery, literally.} The three
operator letters form an inductive type acting on sets of functions, words are lists of
letters acting by composition (rightmost letter first, so that
$[A_+,A_\wedge,A_\wedge](P)=A_+(A_\wedge(A_\wedge(P)))$), and $\F_{\mathrm{Natural}}$ is
the union over all words:
\begin{lstlisting}
inductive OpLetter : Type
  | plus | mul | elev

def OpLetter.apply : OpLetter → Set (ℕ+ → ℕ+) → Set (ℕ+ → ℕ+)
  | plus, P => P ∪ {f | ∃ g ∈ P, ∃ h ∈ P, f = fun n => g n + h n}
  | mul,  P => P ∪ {f | ∃ g ∈ P, ∃ h ∈ P, f = fun n => g n * h n}
  | elev, P => P ∪ {f | ∃ g ∈ P, ∃ h ∈ P, f = fun n => g n ^ (h n : ℕ)}

abbrev Word := List OpLetter

def Word.apply (σ : Word) (P : Set (ℕ+ → ℕ+)) : Set (ℕ+ → ℕ+) :=
  σ.foldr (fun l Q => l.apply Q) P

def Fs : Set (ℕ+ → ℕ+) := {f | f = (fun n => n) ∨ ∃ c : ℕ+, f = fun _ => c}

def FNatural : Set (ℕ+ → ℕ+) := ⋃ σ : Word, Word.apply σ Fs
\end{lstlisting}
Two elementary properties of the operators drive everything that follows: each is
\emph{cumulative} ($P\subseteq A(P)$) and \emph{monotone} ($P\subseteq Q$ implies
$A(P)\subseteq A(Q)$), hence so is every word:
\begin{lstlisting}
theorem OpLetter.subset_apply (l : OpLetter) (P : Set (ℕ+ → ℕ+)) : P ⊆ l.apply P
theorem OpLetter.apply_mono (l : OpLetter) (h : P ⊆ Q) : l.apply P ⊆ l.apply Q
theorem Word.subset_apply (σ : Word) (P : Set (ℕ+ → ℕ+)) : P ⊆ Word.apply σ P
theorem Word.apply_mono (σ : Word) (h : P ⊆ Q) : Word.apply σ P ⊆ Word.apply σ Q
theorem Word.apply_append (σ τ : Word) (P : Set (ℕ+ → ℕ+)) :
    Word.apply (σ ++ τ) P = Word.apply σ (Word.apply τ P)
\end{lstlisting}

\medskip
\noindent\textbf{Formalization, second encoding: the inductive closure.}
$\F_{\mathrm{Natural}}$ is exactly the closure of $\{\mathrm{id}\}\cup\{\text{constants}\}$
under pointwise $+,\times,\wedge$, and that closure has a direct Lean encoding as an
inductive predicate:
\begin{lstlisting}
inductive IsNatural : (ℕ+ → ℕ+) → Prop
  | id : IsNatural (fun n => n)
  | const (c : ℕ+) : IsNatural (fun _ => c)
  | add {f g} (hf : IsNatural f) (hg : IsNatural g) : IsNatural (fun n => f n + g n)
  | mul {f g} (hf : IsNatural f) (hg : IsNatural g) : IsNatural (fun n => f n * g n)
  | elev {f g} (hf : IsNatural f) (hg : IsNatural g) :
      IsNatural (fun n => (f n) ^ (g n : ℕ))
\end{lstlisting}
This encoding hands us, for free, the induction principle the paper obtains from
\cite[Definition~1.7]{Mayeux}: to prove a property of all natural functions, prove it for the identity
and the constants and propagate it through the three operations. It is the form used in
every proof in the file. The two encodings are \emph{proved} to agree:
\begin{lstlisting}
theorem isNatural_of_mem_word {σ : Word} {f : ℕ+ → ℕ+} :
    f ∈ Word.apply σ Fs → IsNatural f

theorem mem_FNatural_of_isNatural {f : ℕ+ → ℕ+} (hf : IsNatural f) : f ∈ FNatural

theorem FNatural_eq : FNatural = {f | IsNatural f}
\end{lstlisting}
The first inclusion is induction on the word $\sigma$: membership in
$\sigma(\F_s)$ either passes to the tail of the word or exposes a top-level pointwise
operation, matching an \lcode{IsNatural} constructor. The proof is short enough to give in
full (the \lcode{mul} and \lcode{elev} cases repeat the \lcode{plus} case):
\begin{lstlisting}
theorem isNatural_of_mem_word : ∀ {σ : Word} {f : ℕ+ → ℕ+},
    f ∈ Word.apply σ Fs → IsNatural f := by
  intro σ
  induction σ with
  | nil =>
    rintro f (rfl | ⟨c, rfl⟩)
    · exact .id
    · exact .const c
  | cons l σ ih =>
    intro f hf
    cases l with
    | plus =>
      rcases hf with hf | ⟨g, hg, k, hk, rfl⟩
      · exact ih hf
      · exact .add (ih hg) (ih hk)
    -- ⋮ (mul, elev: identical, with .mul, .elev)
\end{lstlisting}
The second inclusion is induction on the derivation of \lcode{IsNatural f}: the symbols
are reached by the empty word, and at a binary step with witness words
$\sigma_1,\sigma_2$ for the two arguments, the concatenation
$\sigma_1{+\!\!+}\,\sigma_2$ contains both arguments (cumulativity plus monotonicity,
packaged as \lcode{Word.mem\_\allowbreak apply\_\allowbreak append\_\allowbreak left}
and \lcode{\_right}), so one further letter produces the compound function. The \lcode{add} case, in full:
\begin{lstlisting}
  | @add g k _ _ ihg ihk =>
    obtain ⟨σ₁, h1⟩ := Set.mem_iUnion.1 ihg
    obtain ⟨σ₂, h2⟩ := Set.mem_iUnion.1 ihk
    exact Set.mem_iUnion.2 ⟨OpLetter.plus :: (σ₁ ++ σ₂), Set.mem_union_right _
      ⟨g, Word.mem_apply_append_left h1, k, Word.mem_apply_append_right h2, rfl⟩⟩
\end{lstlisting}
This equivalence discharges the obligation that the file's working definition of
``natural'' is the printed one.

\medskip
\noindent\textbf{Length.} \cite[Definition~1.7]{Mayeux} is formalized on the word side, together with the
fact that on natural functions the minimum is attained (so the definition is well posed):
\begin{lstlisting}
noncomputable def natLength (f : ℕ+ → ℕ+) : ℕ :=
  sInf {n | ∃ σ : Word, σ.length = n ∧ f ∈ Word.apply σ Fs}

theorem natLength_spec {f : ℕ+ → ℕ+} (hf : IsNatural f) :
    ∃ σ : Word, σ.length = natLength f ∧ f ∈ Word.apply σ Fs
\end{lstlisting}
The file's proofs use structural induction on \lcode{IsNatural} derivations where the paper
uses induction on length; the two induction principles reach the same statements, and
\lcode{natLength\_spec} is exactly what connects them.

\medskip
\noindent\textbf{$\F_{\mathrm{Natural}}$ is an elevation structure.} The addendum to
\cite[Definition~1.4]{Mayeux} is formalized as an instance on the subtype of natural functions, with the
operations inherited pointwise and closure provided by the \lcode{IsNatural} constructors:
\begin{lstlisting}
instance : ElevationStructure {f : ℕ+ → ℕ+ // IsNatural f} where
  add f g := ⟨fun n => f.1 n + g.1 n, f.2.add g.2⟩
  mul f g := ⟨fun n => f.1 n * g.1 n, f.2.mul g.2⟩
  elev f g := ⟨fun n => f.1 n ^ (g.1 n : ℕ), f.2.elev g.2⟩
  -- ⋮ (axioms by Subtype.ext from the pointwise instance)
\end{lstlisting}

\section{\texorpdfstring{\cite[Proposition~1.6, Lemmas~1.8--1.10, Remark~1.11]{Mayeux}}{[3, Proposition 1.6, Lemmas 1.8-1.10, Remark 1.11]}}

\noindent\textbf{Proposition~1.6 (\cite{Mayeux}).} \emph{A natural function is constant or
strictly increasing.}

The printed proof is by induction on length, via three closure lemmas: \cite[Lemma~1.8]{Mayeux} (both
arguments strictly increasing $\Rightarrow$ so are their sum, product, and elevation),
\cite[Lemma~1.9]{Mayeux} (one argument strictly increasing, the other constant $\Rightarrow$ sum, product,
and elevation with the constant as exponent are strictly increasing), and \cite[Lemma~1.10]{Mayeux}
(elevation with a \emph{constant base} $c$ and strictly increasing exponent is strictly
increasing \emph{unless $c=1$}, in which case it is constant). This last case split is easy
to miss and is the only place where the argument is not symmetric in its two operands.

\medskip
\noindent\textbf{Formalization.} The proposition is proved by structural induction on the
derivation, reproducing exactly this case analysis, with the \lcode{elev} case carrying the
base-$1$ exception of \cite[Lemma~1.10]{Mayeux} verbatim:
\begin{lstlisting}
def ConstOrStrictMono (f : ℕ+ → ℕ+) : Prop := (∃ c, ∀ n, f n = c) ∨ StrictMono f

theorem isNatural_constOrStrictMono {f : ℕ+ → ℕ+} (hf : IsNatural f) :
    ConstOrStrictMono f
\end{lstlisting}
The three lemmas also exist as standalone named statements, so that the printed proof's
skeleton is visible in the file rather than folded silently into one induction. (The
printed lemmas hypothesize $h,g\in\F_{\mathrm{Natural}}$; the Lean statements take
arbitrary functions, since the printed proofs use only strict monotonicity and constancy;
naturality never enters.)
\begin{lstlisting}
theorem lemma_1_8 {h g : ℕ+ → ℕ+} (hh : StrictMono h) (hg : StrictMono g) :
    StrictMono (fun n => h n + g n) ∧ StrictMono (fun n => h n * g n) ∧
      StrictMono (fun n => h n ^ (g n : ℕ))

theorem lemma_1_9 {h : ℕ+ → ℕ+} (c : ℕ+) (hh : StrictMono h) :
    StrictMono (fun n => h n + c) ∧ StrictMono (fun n => h n * c) ∧
      StrictMono (fun n => h n ^ (c : ℕ))

theorem lemma_1_10 {h : ℕ+ → ℕ+} (c : ℕ+) (hh : StrictMono h) :
    (1 < c → StrictMono (fun n => c ^ (h n : ℕ))) ∧
      (c = 1 → ∀ n, c ^ (h n : ℕ) = 1)
\end{lstlisting}
The elevation part of \cite[Lemma~1.8]{Mayeux} is the one genuinely two-step estimate:
$h(i)^{g(i)} \le h(i)^{g(j)} < h(j)^{g(j)}$ for $i<j$, monotonicity in the exponent
followed by strict monotonicity in the base. In the file this is a two-line \lcode{calc},
on Mathlib's \lcode{pow\_le\_pow\_right'} and \lcode{pow\_lt\_pow\_left'}:
\begin{lstlisting}
    calc h i ^ (g i : ℕ) ≤ h i ^ (g j : ℕ) :=
          pow_le_pow_right' (one_le (a := h i)) (by exact_mod_cast (hg hij).le)
      _ < h j ^ (g j : ℕ) := pow_lt_pow_left' (g j).2.ne' (hh hij)
\end{lstlisting}

\medskip
\noindent\textbf{Remark~1.11 (\cite{Mayeux}).} On $\mathbb{N}$, the map $n\mapsto n^n$ is
neither constant nor strictly increasing (since $0^0=1=1^1$ while $2^2=4$); its restriction
to $\I$ is natural and strictly increasing. This is why the theory's domain is $\I$ and not
$\mathbb{N}$. Both halves are formalized:
\begin{lstlisting}
theorem remark_1_11_not_strictMono : ¬ StrictMono (fun n : ℕ => n ^ n)
theorem remark_1_11_not_const : ¬ ∃ c, ∀ n : ℕ, n ^ n = c
theorem remark_1_11_isNatural : IsNatural (fun n : ℕ+ => n ^ (n : ℕ))
theorem remark_1_11_strictMono : StrictMono (fun n : ℕ+ => n ^ (n : ℕ))
\end{lstlisting}
The first uses precisely the paper's counterexample $0^0=1^1$; the last is \cite[Proposition~1.6]{Mayeux}
applied to \lcode{remark\_1\_11\_isNatural} ($=\lcode{elev id id}$), with constancy ruled
out at $n=1,2$.

\section{\texorpdfstring{\cite[Example~1.5]{Mayeux}}{[3, Example 1.5]}: polynomials, Fermat's function, and the sample functions}
\label{sec:ex15}

\noindent\textbf{Example~1.5 (\cite{Mayeux}).} (a) $\F_s\subseteq\F_{\mathrm{Natural}}$.
(b) The set $\F_{\mathrm{Polynomial}}$ of polynomial functions with positive integer
coefficients satisfies
$\F_{\mathrm{Polynomial}}\subseteq\bigcup_{j,k\ge0}A_+^{\,j}A_\times^{\,k}(\F_s)
\subseteq\F_{\mathrm{Natural}}$.
(c) Fermat's function $n\mapsto2^{2^n}+1$ belongs to $A_+A_\wedge A_\wedge(\F_s)$.
(d) Sample natural functions include
\[
n^n+n+1,\qquad 7^n+6,\qquad n^{\,4n^n+n^{23}+2^n+8}+n+3,\qquad
2^{2^{2^{2^{2^{2^n}}}}}+1.
\]

\medskip
\noindent\textbf{Formalization of (a) and (b).} Part (a) is the empty word:
\begin{lstlisting}
theorem Fs_subset_FNatural : Fs ⊆ FNatural
\end{lstlisting}
For (b), a polynomial function is recorded through its values, since \lcode{ℕ+} has no zero,
so the defining identity is stated in \lcode{ℕ} after coercion, with a nonempty (finite)
support and positive coefficients, exactly the paper's class:
\begin{lstlisting}
def FPolynomial : Set (ℕ+ → ℕ+) :=
  {f | ∃ s : Finset ℕ, s.Nonempty ∧ ∃ c : ℕ → ℕ+,
    ∀ n : ℕ+, (f n : ℕ) = ∑ k ∈ s, (c k : ℕ) * (n : ℕ) ^ k}
\end{lstlisting}
The paper's refined containment is proved with the words exhibited explicitly, as
\lcode{List.replicate}:
\begin{lstlisting}
theorem monomial_mem_mul_word (c : ℕ+) (k : ℕ) :
    (fun n : ℕ+ => c * n ^ k) ∈ Word.apply (List.replicate k OpLetter.mul) Fs

theorem FPolynomial_subset_words :
    ∀ f ∈ FPolynomial, ∃ j k : ℕ,
      f ∈ Word.apply (List.replicate j OpLetter.plus ++ List.replicate k OpLetter.mul) Fs

theorem FPolynomial_subset_FNatural : FPolynomial ⊆ FNatural
\end{lstlisting}
The proof is the expected double induction, and its bookkeeping is exactly where the word
formalism earns its cumulativity and monotonicity lemmas: a monomial $c\cdot n^k$ enters
after $k$ letters $A_\times$ (induction on $k$); for a sum of monomials (induction on the
support, via \lcode{Finset.Nonempty.cons\_induction}), the two summands' witness words are
first lifted to a common word, a \lcode{List.replicate} of the maximum of the two
$A_\times$-counts (using $A^{K}(P)=A^{K-k}(A^{k}(P))\supseteq A^k(P)$), and one further
letter $A_+$ produces the sum. The exponents $j,k$ produced are as in the paper: (number of
monomials $-1$ successive $A_+$'s, after enough $A_\times$'s for the largest monomial).

\medskip
\noindent\textbf{Formalization of (c).} Fermat's function is already central in \S2--\S3
(see \S\ref{sec:prop22} below); here it is placed in the paper's \emph{literal} word,
$[A_+,A_\wedge,A_\wedge]$ acting as $A_+(A_\wedge(A_\wedge(\F_s)))$:
\begin{lstlisting}
def fermatFn : ℕ+ → ℕ+ := fun n => (2 : ℕ+) ^ (((2 : ℕ+) ^ (n : ℕ) : ℕ+) : ℕ) + 1

theorem fermatFn_mem_word :
    fermatFn ∈ Word.apply [OpLetter.plus, OpLetter.elev, OpLetter.elev] Fs
\end{lstlisting}
The proof builds the three levels by hand: $2^n\in A_\wedge(\F_s)$ from the constant $2$
and the identity; $2^{2^n}\in A_\wedge^2(\F_s)$ from the (cumulated) constant $2$ and the
level-one function; then one $A_+$ with the constant $1$.

\medskip
\noindent\textbf{Formalization of (d).} Each sample is a theorem whose proof is nothing but
the explicit generating derivation: the constructor term \emph{is} the paper's
parenthesized expression:
\begin{lstlisting}
theorem example_1_5_selfPow : IsNatural (fun n : ℕ+ => n ^ (n : ℕ) + n + 1) :=
  .add (.add (.elev .id .id) .id) (.const 1)

theorem example_1_5_sevenPow : IsNatural (fun n : ℕ+ => (7 : ℕ+) ^ (n : ℕ) + 6) :=
  .add (.elev (.const 7) .id) (.const 6)

theorem example_1_5_big :
    IsNatural (fun n : ℕ+ =>
      n ^ ((4 * n ^ (n : ℕ) + n ^ ((23 : ℕ+) : ℕ) + (2 : ℕ+) ^ (n : ℕ) + 8 : ℕ+) : ℕ)
        + n + 3) :=
  .add (.add (.elev .id
    (.add (.add (.add (.mul (.const 4) (.elev .id .id)) (.elev .id (.const 23)))
      (.elev (.const 2) .id)) (.const 8))) .id) (.const 3)

theorem example_1_5_tower :
    IsNatural (fun n : ℕ+ =>
      (2 : ℕ+) ^ (((2 : ℕ+) ^ (((2 : ℕ+) ^ (((2 : ℕ+) ^ (((2 : ℕ+) ^
        (((2 : ℕ+) ^ (n : ℕ) : ℕ+) : ℕ) : ℕ+) : ℕ) : ℕ+) : ℕ) : ℕ+) : ℕ) : ℕ+) : ℕ)
        + 1) :=
  .add (.elev (.const 2) (.elev (.const 2) (.elev (.const 2) (.elev (.const 2)
    (.elev (.const 2) (.elev (.const 2) .id)))))) (.const 1)
\end{lstlisting}
The tower's iterated coercions \lcode{((... : ℕ+) : ℕ)} are the formal cost of $\wedge$
associating to the right on \lcode{ℕ+} while Lean's exponent lives in \lcode{ℕ}; the
right-nested \lcode{.elev} chain is exactly the paper's convention
$a_1\wedge(a_2\wedge(\cdots))$ made explicit.

\section{Conjecture 2.1 and the supernatural reformulation (\texorpdfstring{\cite[\S2]{Mayeux}}{[3, Sec. 2]})}

\noindent\textbf{Conjecture~2.1 (\cite{Mayeux}).} \emph{Let
$f\in\F_{\mathrm{Natural}}$ be non-constant. Then $f(\I)\not\subset\mathbb{P}$.}

This is the paper's central conjecture, formulated in 2013. It is stated in Lean exactly as
above, as a named proposition, with \emph{no proof attached}:
\begin{lstlisting}
def Conjecture_2_1 : Prop :=
  ∀ f : ℕ+ → ℕ+, IsNatural f → ¬(∃ c, ∀ n, f n = c) → ∃ n : ℕ+, ¬ Nat.Prime (f n : ℕ)
\end{lstlisting}

\begin{honesty}
This is deliberately a plain \lcode{def : Prop}, not a \lcode{theorem} closed by
\lcode{sorry} and not an \lcode{axiom}: restating the conjecture in Lean changes nothing
about its status. Every result below that uses \lcode{Conjecture\_2\_1} takes it as an
explicit hypothesis, so the dependency structure is visible in each statement.
\end{honesty}

\cite[\S2.2]{Mayeux} reformulates the conjecture set-theoretically: an \emph{infinite} subset
of $\I$ is \emph{natural} if it is $f(\I)$ for some non-constant natural $f$; an
\emph{infinite} subset is \emph{supernatural} if it contains no natural subset;
\cite[Conjecture~2.1]{Mayeux} is then equivalent to ``$\mathbb{P}$ is supernatural''. The Lean
definitions omit the two ``infinite'' qualifiers, which are immaterial: for natural sets
infiniteness is automatic (the range of a strictly increasing function,
\cite[Proposition~1.6]{Mayeux}) and is proved as a theorem; for the supernatural side the
equivalence is applied only to $\mathbb{P}$, which is infinite.
\begin{lstlisting}
def IsNaturalSet (S : Set ℕ+) : Prop :=
  ∃ f : ℕ+ → ℕ+, IsNatural f ∧ ¬(∃ c, ∀ n, f n = c) ∧ S = Set.range f

theorem IsNaturalSet.infinite {S : Set ℕ+} (hS : IsNaturalSet S) : S.Infinite

def IsSupernatural (S : Set ℕ+) : Prop :=
  ∀ T : Set ℕ+, IsNaturalSet T → ¬(T ⊆ S)

theorem conjecture_2_1_iff_supernatural :
    Conjecture_2_1 ↔ IsSupernatural {p : ℕ+ | Nat.Prime (p : ℕ)}
\end{lstlisting}
Both directions of the equivalence are exactly what they should be: an unfolding of
quantifiers once \lcode{IsNaturalSet} is unfolded to expose the underlying natural
function, with no further mathematical content beyond \cite[Conjecture~2.1]{Mayeux} itself.

\section{\texorpdfstring{\cite[Proposition~2.2]{Mayeux}}{[3, Proposition 2.2]}: the proved cases}\label{sec:prop22}

\noindent\textbf{Proposition~2.2 (\cite{Mayeux}).} \emph{Conjecture~2.1 holds in the
following cases:}
\begin{itemize}
\item[(i)] $f$ a non-constant polynomial function on $\I$;
\item[(ii)] $f(n) = a^n+b$ for $a\in\I_{>1}$, $b\in\mathbb{N}$;
\item[(iii)] $f$ Fermat's function $n\mapsto 2^{2^n}+1$;
\item[(iv)] the natural functions tested experimentally in the paper's \S4.
\end{itemize}

\medskip
\noindent\textbf{Case (iii).} Euler's factorization of the fifth Fermat number, checked by
the kernel:
\begin{lstlisting}
theorem fermatFn_isNatural : IsNatural fermatFn
theorem fermatFn_five_eq : (fermatFn 5 : ℕ) = 641 * 6700417 := by decide
theorem fermatFn_five_not_prime : ¬ Nat.Prime (fermatFn 5 : ℕ)
theorem fermatFn_witnesses_conjecture : ∃ n : ℕ+, ¬ Nat.Prime (fermatFn n : ℕ) :=
  ⟨5, fermatFn_five_not_prime⟩
\end{lstlisting}
\lcode{fermatFn} is built from the literal \lcode{IsNatural} constructors (and placed in
its generating word in \S\ref{sec:ex15}), so its naturality is checked, not assumed; the
theorem \lcode{fermatFn\_five\_eq} reproduces the Fermat row of \cite[Table~1]{Mayeux}
($f(n)=2^{2^n}+1$: smallest bad $n$ is $5$, factorization $641\times6700417$).

\medskip
\noindent\textbf{Cases (i) and (ii)}, reproducing the paper's own proofs:
\begin{lstlisting}
theorem prop_2_2_i {f : ℕ+ → ℕ+} (hf : IsNatural f) (q : Polynomial ℤ)
    (hq : ∀ n : ℕ+, (f n : ℤ) = q.eval (n : ℤ)) (hnc : ¬(∃ c, ∀ n, f n = c)) :
    ∃ n : ℕ+, ¬ Nat.Prime (f n : ℕ)

theorem prop_2_2_ii {a : ℕ} (ha : 1 < a) (b : ℕ) :
    ∃ n : ℕ+, ¬ Nat.Prime (a ^ (n : ℕ) + b)
\end{lstlisting}
For (i): by \cite[Proposition~1.6]{Mayeux}, $f$ is strictly increasing, so $f(n)>1$ for some $n$; take a
prime $p\mid f(n)$. The classical fact that $x-y \mid q(x)-q(y)$ for an integer polynomial
$q$ (Mathlib's \lcode{Polynomial.sub\_dvd\_eval\_sub}) gives $f(n+p)\equiv f(n)\equiv0
\pmod p$ directly, with $1<f(n)<f(n+p)$, so $p$ is a proper divisor of $f(n+p)$. For (ii):
take a prime $p\mid f(2)=a^2+b$; if $p\nmid a$, Fermat's little theorem (Mathlib's
\lcode{Nat.ModEq.pow\_totient} specialized to a prime modulus) gives $a^{2+(p-1)}\equiv
a^2\pmod p$, so $p\mid f(2+p-1)$ too, with $1<f(2)<f(2+p-1)$; if $p\mid a$ then already
$p<f(2)$ and $p \mid f(2)$. In Lean, Fermat's little theorem enters as Mathlib's
\lcode{Nat.ModEq.pow\_totient} with \lcode{Nat.totient\_prime}. One inessential
specialization: the printed proof runs at a generic $n\in\I_{>1}$; the Lean proof
instantiates $n=2$, which suffices for the existential conclusion.

One deliberate generalization in (i): rather than restrict to the class
$\F_{\mathrm{Polynomial}}$ of \S\ref{sec:ex15}, \lcode{prop\_2\_2\_i} takes an arbitrary
\lcode{Polynomial ℤ} agreeing with $f$ pointwise, together with \lcode{IsNatural f}
supplying monotonicity via \cite[Proposition~1.6]{Mayeux} (needed since a general integer polynomial,
unlike one with positive coefficients, need not be monotone on its own). Every
$f\in\F_{\mathrm{Polynomial}}$ satisfies the hypotheses, so the printed case (i) is an
instance.

\medskip
\noindent\textbf{Case (iv)} is the subject of the experimental tables, formalized row by
row in \S\ref{sec:tables} below.

\section{Consequences of the conjecture (\texorpdfstring{\cite[\S3]{Mayeux}}{[3, Sec. 3]}) and the extension of Remark 3.4}

\noindent\textbf{Proposition~3.1 (\cite{Mayeux}).} \emph{Assume Conjecture~2.1. Let
$f\in\F_{\mathrm{Natural}}$ be non-constant. Then $\{x\in f(\I) : x\text{ not prime}\}$
is infinite.}

The printed proof is short: by \cite[Proposition~1.6]{Mayeux}, $f$ is strictly increasing, so it suffices
to show that for every $k\in\I$ there is $q>k$ with $f(q)$ not prime; given $k$, ``the
function $g(n):=f(n+k)$ is natural'', so \cite[Conjecture~2.1]{Mayeux} applied to $g$ gives $d$ with
$g(d)$ not prime, and $q:=d+k$ works. The claim that $g$ is natural is asserted without
further comment.

\medskip
\noindent\textbf{Formalization.} Making that one-line assertion precise is exactly where an
auxiliary lemma is needed: \emph{natural functions are closed under composition}, together
with the fact that the shift $n\mapsto n+k$ is itself natural
($=\mathrm{id}+\mathrm{const}\;k$):
\begin{lstlisting}
theorem IsNatural.comp {f g : ℕ+ → ℕ+} (hf : IsNatural f) (hg : IsNatural g) :
    IsNatural (f ∘ g)

theorem isNatural_shift (k : ℕ+) : IsNatural (fun n : ℕ+ => n + k) :=
  IsNatural.add IsNatural.id (IsNatural.const k)
\end{lstlisting}
\lcode{IsNatural.comp} is proved by induction on the derivation of \lcode{f}, carrying
\lcode{IsNatural g} along as a fixed side hypothesis: at each closure step, composing with
\lcode{g} on the right commutes with the pointwise operation, so each case is one
constructor application. With this in hand, \cite[Proposition~3.1]{Mayeux} is proved by contradiction, rather than the printed
proof's direct ``for every $k$'' argument, but resting on the same core step. If
$S=\{n : \neg\mathrm{Prime}(f\,n)\}$ were finite it would be bounded above by some $N$;
the shift-composite $g:=f\circ({\cdot}+N)$ is natural (\lcode{IsNatural.comp}) and strictly
increasing, hence non-constant, and $g(n)=f(n+N)$ is prime for every $n$ (as $n+N>N$ lies
outside $S$): a non-constant natural function with \emph{no} non-prime value at all,
contradicting \cite[Conjecture~2.1]{Mayeux} outright.
\begin{lstlisting}
def Proposition_3_1 : Prop :=
  Conjecture_2_1 →
    ∀ f : ℕ+ → ℕ+, IsNatural f → ¬(∃ c, ∀ n, f n = c) →
      {n : ℕ+ | ¬ Nat.Prime (f n : ℕ)}.Infinite

theorem proposition_3_1 : Proposition_3_1
\end{lstlisting}

The statement just proved quantifies over the \emph{index} set
$\{n:\neg\mathrm{Prime}(f\,n)\}$; the printed statement quantifies over the \emph{value}
set $\{x\in f(\I):x\text{ not prime}\}$. Since $f$ is injective (strictly increasing), the
two sets are in bijection via $f$, and the printed form is formalized as well, as the image
of the first under $f$:
\begin{lstlisting}
theorem proposition_3_1_values (hconj : Conjecture_2_1) {f : ℕ+ → ℕ+}
    (hf : IsNatural f) (hnc : ¬(∃ c, ∀ n, f n = c)) :
    {x : ℕ+ | x ∈ Set.range f ∧ ¬ Nat.Prime (x : ℕ)}.Infinite
\end{lstlisting}

\medskip
\noindent\textbf{Corollary~3.2 (\cite{Mayeux}).} \emph{Assume Conjecture~2.1. Then there
are infinitely many composite Fermat numbers.} Proved by specializing \cite[Proposition~3.1]{Mayeux}:
\begin{lstlisting}
theorem corollary_3_2 (hconj : Conjecture_2_1) :
    {n : ℕ+ | ¬ Nat.Prime (fermatFn n : ℕ)}.Infinite :=
  proposition_3_1 hconj fermatFn fermatFn_isNatural fermatFn_not_const
\end{lstlisting}
As the paper stresses, this would settle, conditionally on \cite[Conjecture~2.1]{Mayeux}, a question that
is open unconditionally today.

\medskip
\noindent\textbf{Proposition~3.3 (\cite{Mayeux}).} \emph{Assume Conjecture~2.1 is wrong.
Then there exists an ``arithmetical'' (relying only on $+,\times,\wedge$) formula giving
arbitrary big prime numbers.} The printed proof exhibits the content behind that
phrasing: a non-constant natural function (hence, by \cite[Proposition~1.6]{Mayeux}, strictly increasing,
so taking arbitrarily large values) all of whose values are prime. The Lean
theorem records exactly that content, as an immediate unfolding of
$\neg\mathrm{Conjecture\_2\_1}$:
\begin{lstlisting}
theorem exists_all_prime_of_not_conjecture (h : ¬ Conjecture_2_1) :
    ∃ f : ℕ+ → ℕ+, IsNatural f ∧ StrictMono f ∧ ∀ n, Nat.Prime (f n : ℕ)
\end{lstlisting}

\medskip
\noindent\textbf{Remark~3.4 (\cite{Mayeux}): the extended class.} The paper proposes
enlarging the generating symbols beyond $+,\times,\wedge$, mentioning Knuth's arrows
$\uparrow^j$ (where $\uparrow^1=\wedge$) and the factorial. The formalization implements
one extension combining three new features at once: Knuth arrows whose level is itself a
function in the class, the factorial, and truncated subtraction, admitted only under the
pointwise hypothesis $g(n)<f(n)$, which keeps values in $\I$. Mathlib's
\lcode{hyperoperation} indexes the full hierarchy from zero: index~$0$ is the successor
$b\mapsto b+1$, index~$1$ is addition $a+b$, index~$2$ is multiplication $a\cdot b$,
index~$3$ is exponentiation $a^{b}$, that is, the paper's elevation $\wedge$, index~$4$
is tetration, and so on. Knuth's arrows begin at exponentiation, so $j$ arrows
correspond to index $j+2$:
\[
a\uparrow^{j}b \;=\; \lcode{hyperoperation}\;(j+2)\;a\;b \qquad (j\ge1),\qquad
\uparrow^{1}=\wedge.
\]
This is why the \lcode{knuth} constructor below uses the index $(g\,n)+2$, and why
indices $0$--$2$ are not needed: addition and multiplication are constructors of the
class in their own right. A positivity lemma transports the arrows to \lcode{ℕ+}:
\begin{lstlisting}
theorem hyperoperation_pos_of_three_le {m a : ℕ} (hm : 3 ≤ m) (ha : 0 < a) (b : ℕ) :
    0 < hyperoperation m a b

inductive IsNaturalKnuthFactorialSub : (ℕ+ → ℕ+) → Prop
  | id : IsNaturalKnuthFactorialSub (fun n => n)
  | const (c : ℕ+) : IsNaturalKnuthFactorialSub (fun _ => c)
  | add {f g} (hf : IsNaturalKnuthFactorialSub f) (hg : IsNaturalKnuthFactorialSub g) :
      IsNaturalKnuthFactorialSub (fun n => f n + g n)
  | mul {f g} (hf : IsNaturalKnuthFactorialSub f) (hg : IsNaturalKnuthFactorialSub g) :
      IsNaturalKnuthFactorialSub (fun n => f n * g n)
  | knuth {f g h} (hf : IsNaturalKnuthFactorialSub f) (hg : IsNaturalKnuthFactorialSub g)
      (hh : IsNaturalKnuthFactorialSub h) :
      IsNaturalKnuthFactorialSub (fun n =>
        ⟨hyperoperation ((g n : ℕ) + 2) (f n) (h n),
          hyperoperation_pos_of_three_le (by have := (g n).pos; omega) (f n).pos (h n)⟩)
  | fact {f} (hf : IsNaturalKnuthFactorialSub f) :
      IsNaturalKnuthFactorialSub (fun n => ⟨(f n : ℕ).factorial, (f n : ℕ).factorial_pos⟩)
  | sub {f g} (hf : IsNaturalKnuthFactorialSub f) (hg : IsNaturalKnuthFactorialSub g)
      (hlt : ∀ n, (g n : ℕ) < (f n : ℕ)) :
      IsNaturalKnuthFactorialSub (fun n =>
        ⟨(f n : ℕ) - (g n : ℕ), Nat.sub_pos_of_lt (hlt n)⟩)

theorem IsNatural.toKnuthFactorialSub {f : ℕ+ → ℕ+} (hf : IsNatural f) :
    IsNaturalKnuthFactorialSub f
\end{lstlisting}
The \lcode{knuth} constructor takes \emph{three} functions $f,g,h$ of the class and
produces $n\mapsto f(n)\uparrow^{g(n)}h(n)$: the arrow level varies with the argument.
The embedding \lcode{IsNatural.toKnuthFactorialSub} verifies that the extension does
extend: $\wedge$ is the arrow of constant level~$1$ (via \lcode{hyperoperation\_three}).

The non-triviality hypothesis of the extended conjecture is phrased as the infinitude of
the value set; for plain natural functions this is equivalent to the non-constancy
hypothesis of \cite[Conjecture~2.1]{Mayeux} (\cite[Proposition~1.6]{Mayeux}):
\begin{lstlisting}
def Conjecture_KnuthFactorialSub : Prop :=
  ∀ f : ℕ+ → ℕ+, IsNaturalKnuthFactorialSub f → (Set.range f).Infinite →
    ∃ n : ℕ+, ¬ Nat.Prime (f n : ℕ)
\end{lstlisting}
The extended conjecture is neither proved nor assumed, matching the printed remark's
status exactly.

\section{The experimental tables (\texorpdfstring{\cite[\S4]{Mayeux}}{[3, Sec. 4]})}\label{sec:tables}

\cite[\S4]{Mayeux} reports three tables of experiments, each row giving a natural function,
the smallest $n$ for which its value is not prime, and that value's factorization. This is
the content of \cite[Proposition~2.2(iv)]{Mayeux}, and it is formalized exhaustively: \emph{every row of
every table} is a theorem, stated as an explicit conjunction (prime at each earlier
point, not prime at the reported one) and proved by \lcode{norm\_num}
\cite{mathlib}, except for the seven largest primality conjuncts, which are proved by
kernel-checked Lucas certificates (below). \cite[Table~1]{Mayeux} collects ten assorted non-polynomial functions; \cite[Table~2]{Mayeux} treats
$f_k(n)=2^{2^n}+2k+1$ for $k=1,\dots,59$; \cite[Table~3]{Mayeux} treats $n\mapsto2^{2^n}+c$ for the
twenty odd $c\in\{2501,2503,\dots,2539\}$. Representative rows:
\begin{lstlisting}
theorem table1_row1 :
    Nat.Prime ((3:ℕ) ^ 1 + 2 ^ (1 + 1)) ∧ Nat.Prime ((3:ℕ) ^ 2 + 2 ^ (2 + 1)) ∧
    Nat.Prime ((3:ℕ) ^ 3 + 2 ^ (3 + 1)) ∧ Nat.Prime ((3:ℕ) ^ 4 + 2 ^ (4 + 1)) ∧
    Nat.Prime ((3:ℕ) ^ 5 + 2 ^ (5 + 1)) ∧ Nat.Prime ((3:ℕ) ^ 6 + 2 ^ (6 + 1)) ∧
    ¬ Nat.Prime ((3:ℕ) ^ 7 + 2 ^ (7 + 1)) := by
  norm_num

theorem table3_row2501 :
    ¬ Nat.Prime ((2:ℕ) ^ (2 ^ 1) + 2501) := by
  norm_num
\end{lstlisting}
Two points deserve comment.

First, the encoding choice: rather than re-transcribe the paper's reported factorizations
(huge, and easy to mis-transcribe), each $\neg\lcode{Nat.Prime}$ conjunct is proved by
\lcode{norm\_num} finding its own witness to compositeness: equally rigorous, and immune
to a copying error in a $39$-digit factorization string. The one exception is Euler's
$641\times6700417$, kept as an explicit identity (\lcode{fermatFn\_five\_eq}) for its
historical weight.

Second, trial division does not scale to the largest entries. The paper \cite{Mayeux} highlights
$f_{46}(n)=2^{2^n}+93$ as beating Fermat's own function: its values are prime for
$n=1,\dots,6$, so the corresponding theorem certifies, among its conjuncts, the primality
of $2^{64}+93\approx1.8\times10^{19}$; \cite[Table~3]{Mayeux} contains one row of the same weight
($c=2535$, the paper's second function prime for all $1\le n<7$). A \lcode{norm\_num}
proof of such a conjunct is a kernel-checked trial division with about $2^{31}$ steps,
and these conjuncts alone used to cost about an hour of \lcode{lake build}. They are
instead proved by \emph{Lucas certificates}: to certify $p$ prime it suffices to exhibit
a witness $a$ with $a^{p-1}\equiv1\pmod p$ and $a^{(p-1)/q}\not\equiv1\pmod p$ for each
prime $q\mid p-1$. Mathlib supplies both halves of the work: the test itself is
\lcode{lucas\_primality}, and the modular-exponentiation conditions, stated in
\lcode{ZMod p}, are discharged by the \lcode{reduce\_mod\_char} tactic, whose
\lcode{norm\_num} extension evaluates $a^b \bmod m$ by binary modular exponentiation.
The only project-level glue is one short lemma, in the file's \lcode{Pratt}
namespace, converting an explicit factorization of $p-1$ into
\lcode{lucas\_primality}'s quantification over prime divisors:
\begin{lstlisting}
theorem lucasCert (p a : ℕ) (l : List ℕ)
    (hl : ∀ q ∈ l, Nat.Prime q)
    (hprod : p - 1 = l.prod)
    (h1 : (a : ZMod p) ^ (p - 1) = 1)
    (h2 : ∀ q ∈ l, (a : ZMod p) ^ ((p - 1) / q) ≠ 1) :
    Nat.Prime p
\end{lstlisting}
The certificate data, i.e.\ the factorization of $p-1$ and the witness $a$, is found by
an external computation (seconds in any computer-algebra system) and is \emph{not
trusted}: \lcode{hprod} is checked by \lcode{norm\_num}, \lcode{h1} and \lcode{h2} by
\lcode{reduce\_mod\_char} (with \lcode{decide} closing the resulting numeral
disequalities), and the prime factors listed in \lcode{l} are certified recursively:
sixteen certificates in all, down to primes small enough for \lcode{norm\_num}'s trial
division. The chain culminates in
\begin{lstlisting}
theorem prime_18446744073709551709 : Nat.Prime 18446744073709551709 := by
  refine lucasCert 18446744073709551709 2 [2, 2, 3, 3, 29, 38652541, 457131527]
    ?_ (by norm_num) (by reduce_mod_char) ?_
  · intro q hq
    simp only [List.mem_cons, List.not_mem_nil, or_false] at hq
    rcases hq with rfl | rfl | rfl | rfl | rfl | rfl | rfl
    exacts [by norm_num, by norm_num, by norm_num, by norm_num, by norm_num,
      prime_38652541, prime_457131527]
  · intro q hq
    simp only [List.mem_cons, List.not_mem_nil, or_false] at hq
    rcases hq with rfl | rfl | rfl | rfl | rfl | rfl | rfl <;> (reduce_mod_char; decide)
\end{lstlisting}
and the heavy rows now read
\begin{lstlisting}
theorem table2_row46 :
    Nat.Prime ((2:ℕ) ^ (2 ^ 1) + 2 * 46 + 1) ∧ Nat.Prime ((2:ℕ) ^ (2 ^ 2) + 2 * 46 + 1) ∧
    Nat.Prime ((2:ℕ) ^ (2 ^ 3) + 2 * 46 + 1) ∧ Nat.Prime ((2:ℕ) ^ (2 ^ 4) + 2 * 46 + 1) ∧
    Nat.Prime ((2:ℕ) ^ (2 ^ 5) + 2 * 46 + 1) ∧ Nat.Prime ((2:ℕ) ^ (2 ^ 6) + 2 * 46 + 1) ∧
    ¬ Nat.Prime ((2:ℕ) ^ (2 ^ 7) + 2 * 46 + 1) := by
  refine ⟨by norm_num, by norm_num, by norm_num, by norm_num, ?_, ?_, by norm_num⟩
  · rw [show (2:ℕ) ^ (2 ^ 5) + 2 * 46 + 1 = 4294967389 from by norm_num]
    exact Pratt.prime_4294967389
  · rw [show (2:ℕ) ^ (2 ^ 6) + 2 * 46 + 1 = 18446744073709551709 from by norm_num]
    exact Pratt.prime_18446744073709551709
\end{lstlisting}
The same treatment covers $3^{27}+16$ and $3^{27}+34$ (\cite[Table~1]{Mayeux}, rows~7 and~8) and the
$2^{32}$-sized conjuncts.

\section{Correspondence table}\label{sec:table}

Statuses: \textbf{proved} means stated and proved in Lean with no \lcode{sorry};
\textbf{stated} means recorded as a named \lcode{Prop} with no proof attached,
deliberately, matching its open status in the paper.

\begin{longtable}{@{}p{0.30\textwidth}p{0.40\textwidth}p{0.22\textwidth}@{}}
\toprule
\textbf{\cite{Mayeux}} & \textbf{\lcode{SPCL.lean}} & \textbf{Status}\\
\midrule
\endhead
Def.~1.1 (elevation structure) & \lcode{ElevationStructure} & formalized\\
--- (morphism) & \lcode{ElevationHom} & formalized\\
--- (``we obtain a category'') & \lcode{ElevationHom.id}, \lcode{.comp}, the three laws, and the \lcode{Category ElevCat} instance & \textbf{proved}\\
Ex.~1.2, $\I$ & \lcode{instance : ElevationStructure ℕ+} & formalized\\
Ex.~1.2, $\F$, $E_a$ & \lcode{instance : ElevationStructure (ℕ+ → ℕ+)}, \lcode{Eval} & formalized\\
Def.~1.3 ($A_+,A_\times,A_\wedge$) & \lcode{OpLetter}, \lcode{OpLetter.apply} & formalized\\
Def.~1.4 ($\Sigma$, $\F_{\mathrm{Natural}}$) & \lcode{Word}, \lcode{Word.apply}, \lcode{Fs}, \lcode{FNatural} & formalized\\
--- (inductive counterpart) & \lcode{IsNatural}; equivalence \lcode{FNatural\_eq} & \textbf{proved}\\
--- ($\F_{\mathrm{Natural}}$ elevation structure) & \lcode{instance : ElevationStructure \{f // IsNatural f\}} & formalized\\
Ex.~1.5 (a) $\F_s$ & \lcode{Fs\_subset\_FNatural} & \textbf{proved}\\
Ex.~1.5 (b) polynomials & \lcode{FPolynomial}, \lcode{FPolynomial\_\allowbreak subset\_\allowbreak words}, \lcode{FPolynomial\_\allowbreak subset\_\allowbreak FNatural} & \textbf{proved}\\
Ex.~1.5 (c) Fermat's word & \lcode{fermatFn\_mem\_word} & \textbf{proved}\\
Ex.~1.5 (d) four samples & \lcode{example\_1\_5\_selfPow}, \lcode{\_sevenPow}, \lcode{\_big}, \lcode{\_tower} & \textbf{proved}\\
Prop.~1.6 & \lcode{isNatural\_\allowbreak constOrStrictMono} & \textbf{proved}\\
Def.~1.7 (length) & \lcode{natLength}, \lcode{natLength\_spec} & \textbf{proved}\\
Lemmas 1.8, 1.9, 1.10 & \lcode{lemma\_1\_8}, \lcode{lemma\_1\_9}, \lcode{lemma\_1\_10} & \textbf{proved}\\
Remark 1.11 & \lcode{remark\_1\_11\_*} (four statements) & \textbf{proved}\\
Conjecture~2.1 & \lcode{Conjecture\_2\_1} & \textbf{stated} (open)\\
\S2.2 (supernatural sets) & \lcode{IsNaturalSet}, \lcode{IsSupernatural}, \lcode{conjecture\_\allowbreak 2\_1\_\allowbreak iff\_\allowbreak supernatural}, \lcode{IsNaturalSet.infinite} & \textbf{proved}\\
Prop.~2.2(i) & \lcode{prop\_2\_2\_i} & \textbf{proved}\\
Prop.~2.2(ii) & \lcode{prop\_2\_2\_ii} & \textbf{proved}\\
Prop.~2.2(iii) & \lcode{fermatFn\_five\_eq}, \lcode{fermatFn\_five\_not\_prime} & \textbf{proved}\\
Prop.~2.2(iv) & the $89$ table theorems (\S\ref{sec:tables}) & \textbf{proved}\\
--- & \lcode{IsNatural.comp} & \textbf{proved} (auxiliary for 3.1)\\
Prop.~3.1 & \lcode{proposition\_3\_1}; printed form \lcode{proposition\_3\_1\_values} & \textbf{proved} (conditional)\\
Cor.~3.2 & \lcode{corollary\_3\_2} & \textbf{proved} (conditional)\\
Prop.~3.3 & \lcode{exists\_all\_prime\_\allowbreak of\_not\_conjecture} & \textbf{proved}\\
Remark 3.4 (extended class) & \lcode{IsNaturalKnuth\allowbreak FactorialSub}, \lcode{IsNatural.\allowbreak toKnuthFactorialSub}, \lcode{Conjecture\_\allowbreak KnuthFactorial\allowbreak Sub} & \textbf{stated} (extension), embedding \textbf{proved}\\
\S4, Table 1 & \lcode{table1\_row1} -- \lcode{table1\_row10} & \textbf{proved}\\
\S4, Table 2 ($k=1,\dots,59$) & \lcode{table2\_row1} -- \lcode{table2\_row59} & \textbf{proved}\\
\S4, Table 3 ($c=2501,\dots,2539$) & \lcode{table3\_row2501} -- \lcode{table3\_row2539} & \textbf{proved}\\
--- (primality certificates) & \lcode{Pratt.lucasCert} (over Mathlib's \lcode{lucas\_\allowbreak primality} and \lcode{reduce\_\allowbreak mod\_\allowbreak char}), sixteen \lcode{Pratt.prime\_*} theorems & \textbf{proved}\\
\S4, question (i) (for each $m$, a non-polynomial natural function prime up to $m$) & \lcode{Question\_i} & \textbf{stated} (open)\\
\S4, question (ii) & not a determinate proposition as printed; not encoded & ---\\
\bottomrule
\end{longtable}


\begin{thebibliography}{9}

\bibitem{Lean4} L.~de~Moura and S.~Ullrich, \emph{The Lean 4 theorem prover and
  programming language}, in: Automated Deduction -- CADE~28, Lecture Notes in Computer
  Science, vol.~12699, Springer, 2021, pp.~625--635.

\bibitem{mathlib} The mathlib Community, \emph{The Lean mathematical library}, in:
  Proceedings of the 9th ACM SIGPLAN International Conference on Certified Programs and
  Proofs (CPP 2020), ACM, 2020, pp.~367--381.

\bibitem{Mayeux} A.~Mayeux, \emph{Conjecture: the set of prime numbers is supernatural},
  Proceedings of the Bulgarian Academy of Sciences \textbf{78} (2025), no.~11,
  1585--1592. \href{https://doi.org/10.7546/CRABS.2025.11.01}{doi:10.7546/CRABS.2025.11.01}.

\bibitem{Benchmark} A.~Mayeux, \emph{Formalizing all indexed mathematics as a benchmark
  for general reasoning}, in: Intelligent Systems and Applications: Proceedings of the
  2026 Intelligent Systems Conference (IntelliSys), Lecture Notes in Networks and
  Systems, Springer, to appear. arXiv:2606.03835.

\end{thebibliography}
\end{document}